\documentclass{article}

\usepackage{arxiv}

\usepackage[colorlinks,urlcolor=blue,linkcolor=blue,citecolor=blue]{hyperref}
\usepackage[utf8]{inputenc}

\usepackage{longtable}

\usepackage{color}
\usepackage{lastpage}
\usepackage{enumerate}
\usepackage{mathrsfs}
\usepackage{graphicx}
\usepackage{listings}
\usepackage{array}
\usepackage{amsthm}
\usepackage{amsmath}
\usepackage{amssymb}
\usepackage{bbm}
\usepackage{amscd}
\usepackage{gensymb}
\usepackage{tikz}
\usepackage{pgfplots}
\usepackage{dsfont}
\usepackage{setspace}
\usepackage{tabstackengine}
\usepackage{mathtools}
\usepackage{multirow}
\usepackage{wasysym}
\usepackage{diagbox}
\usepackage{booktabs}
\usepackage[flushleft]{threeparttable}
\usepackage{latexsym}
\usepackage{amsfonts}
\usepackage{amstext}
\usepackage{bm}
\usepackage{cleveref}
\usepackage{pifont}
\usepackage{algorithm}
\usepackage{algorithmic}

\let\SF \mathsf
\let\CAL \mathcal
\let\BBM \mathbbm

\newcommand{\cmark}{\ding{51}}%
\newcommand{\xmark}{\ding{55}}%

\newcommand\norm[1]{\left\lVert#1\right\rVert}
\newcommand\abs[1]{\left\lvert#1\right\rvert}

\newcommand{\beqq}[1]{\begin{align*} #1 \end{align*}}

\newcommand{\transpose}{\mathsf{T}}

\renewcommand{\Re}{\mathbb{R}}
\newcommand\Na{\mathbb{N}}

\newcommand{\ex}[1]{\mathbb{E}\left[#1\right]}
\newcommand\cex[2]{\BBM{E}\left[\left.#1\right\vert#2\right]}
\newcommand\exsub[2]{\BBM{E}_{#1}\left[#2\right]}
\newcommand\pr[1]{\BBM{P}\left(#1\right)}

\newenvironment{renum}
{\begin{enumerate}}
{\end{enumerate}}

\newcommand{\kwh}{\text{ kWh}}
\newcommand{\h}{\text{ h}}

\allowdisplaybreaks

\newtheorem{theorem}{Theorem}
\newtheorem{lemma}{Lemma}

\newtheorem{assumption}{Assumption}

\newtheorem{remark}{Remark}

\title{Preemptive Scheduling of EV Charging for Providing Demand Response Services\thanks{This project is supported through a University Alliance Project from the Ford Motor Company. The contents reported in this research paper are protected by The Ohio State University and Ford Motor Company.}}

\author{\hspace{1mm}Shiping Shao \\
	Department of Electrical and Computer Engineering \\
	The Ohio State University \\
	Columbus, OH, USA. \\
	\texttt{shao.367@osu.edu}
	\And
	\hspace{1mm}Farshad Harirchi \\
	Ford Motor Company \\
	Dearborn, MI, USA.\\
	\texttt{farshad.harirchi@gmail.com} \\
	\AND
	\hspace{1mm}Devang Dave \\
    Ford Motor Company \\
	Dearborn, MI, USA.\\
	\texttt{ddave8@ford.com} \\
	\And
	\hspace{1mm}Abhishek Gupta \\
	Department of Electrical and Computer Engineering \\
	The Ohio State University \\
	Columbus, OH, USA. \\
	\texttt{gupta.706@osu.edu} \\
}

\renewcommand{\shorttitle}{Preemptive Scheduling of EV Charging for Providing Demand Response Services}

\hypersetup{
pdftitle={Preemptive Scheduling of EV Charging for Providing Demand Response Services},
pdfsubject={eess,}
pdfauthor={Shiping Shao},
pdfkeywords={First keyword, Second keyword, More},
}

\begin{document}
\maketitle

\begin{abstract}
We develop a new algorithm for scheduling the charging process of a large number of electric vehicles (EVs) over a finite horizon. We assume that EVs arrive at the charging stations with different charge levels and different flexibility windows. The arrival process is assumed to have a known distribution and that the charging process of EVs can be preemptive. We pose the scheduling problem as a dynamic program with constraints. We show that the resulting formulation leads to a monotone dynamic program with Lipschitz continuous value functions that are robust against perturbation of system parameters. We propose a simulation based fitted value iteration algorithm to determine the value function approximately, and derive the sample complexity for computing the approximately optimal solution. 
\end{abstract}


\section{Introduction}
Passenger electric vehicle sales jumped from 450,000 in 2015 to 2.1 million in 2019. Bloomberg New Energy Finance projects that 8.5 million EVs will be sold every year by 2025 and it will jump to 54 million by 2040 \cite{BNEF2020}. Electric vehicles provide significant advantages with respect to emissions and energy efficiency, and it has the potential to decelerate climate change. EVs also have inherent flexibility in their charging processes, which can be used to provide ancillary services to the grid such as peak shaving and demand response. For instance, EVs at the workplace are parked for over 6-8 hours, but most of them can be fully charged within 1-2 hours based on reasonable usage. Multiple large EV manufacturers have teamed up with several large utility companies in the United States to create a universal platform called Open Vehicle Grid Integration Platform (OVGIP). This platform is expected to aggregate the charging of all EVs manufactured by these automakers and will add an invaluable demand response capability to the grid \cite{OVGIP2014}. Our goal in this paper is to use approximate dynamic programming (ADP) to compute a scheduling algorithm for the charging of EVs at scale. We assume that the charging process can be preemptive, that is, the charging process can be stopped anytime and can be restarted again at a later time, as long as the vehicle is at the desired state of charge (SoC) by the end of the parking time.  

We describe the charging request by an EV as follows: an EV arrives at the charging station with 1) an initial SoC, 2) the target SoC and 3) the departure time. Since the price of electricity and the arrivals of EVs vary over time, we are interested in maximizing the profit of the EV charging aggregation platform by picking an appropriate schedule for charging EVs.

\subsection{Related works}
Scheduling algorithms for EV charging have been developed under various assumptions and goals. For instance, the charging process can be preemptive or it can be non-preemptive. The future demand or its distribution may be known or may be unknown. The goal of the charging process could be to minimize the cost of electricity, to maximize the profit to the load aggregator, or to provide ancillary services to the grid. In this paper, our focus is on devising the EV charging schedule with the assumption of preemptive charging with known demand distribution throughout the day. We outline various strands of research on related topics below:  

\subsubsection{Scheduling with Future Demand Information}
In this case, the charging station has complete information about the future demand, and thus, the charging process can be scheduled offline by a deterministic optimization algorithm with desired objectives. Algorithms proposed in \cite{ma2011decentralized,gan2012optimal,ma2014distributed,sun2016optimal} solved valley-filling problems, which schedule the charging processes to fill the valley in the grid load curve under demand constraints. By assuming a cheaper charging cost when the electricity demand is low, the profit/revenue maximization problem can be formulated similarly as a valley-filling problem. For instance, Wang et al. \cite{wang2018optimal} maximized the revenue of the charging station with time-varying profits, and developed a greedy algorithm to decline demands that are hard to satisfy; Ahn et al. \cite{ahn2011optimal} optimized the joint cost of electricity generation and CO2 emissions, and controlled the load shifting and frequency regulation with a decentralized algorithm; Chen et al. \cite{chen2014electric} maximized the social welfare taking both the power generation cost and the EV charging cost into account. The optimization scheduling can also be obtained through a decentralized algorithm by the convexity of the optimal value function. Besides, the scheduling problem can also be formulated with game theoretic approaches. Mediwaththe et al. \cite{mediwaththe2018game} designed a two-stage non-cooperative game where the charging stations are the players and aim at selecting the optimal charging time to avoid the surge electricity prices. Azimian et al. \cite{azimian2018stackelberg} applied Stackelberg game to study the charging scheduling problem. The charging station is considered a leader that determines the optimal charging time for each EV to shave the peak load according to the best responses from the customers. More recently, Dahlin et al. \cite{dahlin2022scheduling} devised a scheduling algorithm assuming all the demand is known upfront and the customers have been informed about their utility function to the platform; in addition, the paper assumes that once the EV charging starts, it cannot be stopped -- the charging process is non-preemptive.  
    
\subsubsection{Scheduling with Knowledge of Future Demand Distribution}
The knowledge of the distribution of the future demand allows the charging stations to use stochastic optimization methods, i.e., Model Predictive Control (MPC) or scenario-based algorithms to optimize the charging schedule. The problem can be formulated with a mixed integer programming \cite{di2014electric,su2014model} or a rank-constrained optimization problem \cite{shi2018model} to minimize the charging cost. Algorithms using MPC to track a specified demand trajectory are also introduced in \cite{bansal2014plug,Wenzel2018,Wang2020}.

    
The distribution of customers' valuation of the charging service can be used for designing scheduling algorithms with induced demand. Zhang et al. \cite{zhang2018optimal} used a queueing network to model the charging process of EVs. The schedule is optimized by setting a pricing policy to induce the customers to select the appropriate charging mode that minimizes the service drop rate. Liu et al. \cite{liu2018two} introduced a scheduling algorithm when charging demand is elastic and is price sensitive. The authors aim to determine a pricing policy that minimizes the expected operational cost subjected to an energy balance constraint based on expected elastic demand in a day-ahead market.
    
\subsubsection{Scheduling with Instant Demand Information} 
If there is a lack of information about the future demand, then the scheduling problem can be formulated as an online optimization problem. For instance, Chen et al. \cite{chen2012large, chen2012deadline} proposed the Early Deadline First (EDF) based algorithm to schedule the EV charging process and demonstrated that it achieves the highest competitive ratio in comparison with other online algorithms. Gupta et al. \cite{Gupta2015} also applied EDF to schedule the charging process and studied the pricing framework using Lagrange multipliers. Nakahira \cite{Nakahira2017} improved the Least Laxity First (LLF) algorithm by maximizing the minimum laxity of the demands, which yields a higher competitive ratio and less interruption during the charging.

\subsubsection{Scheduling in Data-driven Manner}
More recently, the charging schedule is determined using model-free reinforcement learning algorithms. Wang et al. \cite{wang2019reinforcement} developed a SARSA (state–action–reward–state–action) based algorithm to solve a constrained Markov Decision Problem (MDP) with the charging profits as reward function, where the state–value function is approximated by a linear function. Wan et al. \cite{wan2018model} developed a deep reinforcement learning algorithm to determine the scheduling strategy given the EVs can charge and discharge at the station to shave the peak in grid loads.

There are certain limitations when we implement these algorithms: first, in most scenarios, the distribution of the charging demands can be inferred from the historical dataset. In this case, the online algorithms fail to schedule the charging processes optimally as they do not incorporate the information that can be inferred through the past data into the determination of the charging schedule. In addition, most of the algorithms schedule the charging processes for individual EVs, which means that the computational burden aggravates as the number of EVs increases. Though some of the previous works \cite{tang2016model,Wang2020} aggregate the demands at each time to reduce the dimensionality of the actions in MPC, it still requires the knowledge of the state of each EV in the optimization. Thus, it is difficult to implement these algorithms in a large market with millions of EVs providing grid services. Further, the SARSA based scheduling algorithms require a large number of samples. In this paper, we develop a scheduling algorithm that is sample efficient and scalable, which addresses the problems mentioned above.

\subsection{Our Contribution}
We formulate the problem of preemptive scheduling of charging of a large number of EVs providing grid services as a stochastic dynamic program with a state-dependent action constraint. The key contributions of this paper are as follows:
\begin{enumerate}
    \item We propose a new methodology for assigning each EV that arrives at the charging station a category, which depends on its arrival/departure time and the difference between the target SoC and the initial SoC. The goal of the platform thus is to allocate enough electricity to each category of vehicles so that all the vehicles are charged by their departure time. With this key idea, we identify the state space, the action space, the state transition function, the admissible action set (which depends on the state), and the cost function at each time step. This leads to a stochastic dynamic program.
    \item Under reasonable conditions, the state and the action spaces are very high dimensional (of the order of $\Na^{50}\times\Re^{52}$). We address this problem by using approximate dynamic programming techniques, which use an empirical version of the Bellman operator with function approximation to approximately compute the value function at each step of the decision process. This algorithm is usually referred to as fitted value iteration in the context of stationary infinite-horizon Markov decision problems (our problem has a finite horizon and is non-stationary).
    \item We show the consistency properties of the fitted value iteration using a novel iterated random operator theory, developed in \cite{haskell2016empirical,gupta2015empirical,li2021fitted}. In other words, we show that as (a) the number of samples used in the empirical Bellman operator, (b) the size of the function approximating class, and (c) the number of state samples used for the projecting the approximate value function onto the function approximating class, increase to infinity, the empirical fitted value functions converges to the truly optimal value function. 
    \item We employed comparative statics to show that the value function at each time step satisfies certain monotonicity property in the state and is Lipschitz continuous. We also demonstrate that the value function is continuous and the optimal scheduling policy is lower semi-continuous to the system parameters, like charging costs and EV arrival processes.
    \item We further illustrate our theoretical findings through numerical simulations.
\end{enumerate}

\subsection{Outline of the Paper}
The paper is organized as follows. In Section \ref{sec:scheduling}, we formulate the dynamic scheduling problem and identify the state space, action space, state transition function, and the state-dependent action set. We discuss the key challenges in solving the resulting stochastic dynamic program with state dependent constraints. In Section \ref{sec:efvi}, we introduce our approximate dynamic programming methodology -- fitted value iteration algorithm -- and state our key assumptions and the main result. To prove the main result, we need some auxiliary properties of the value function, which is established in Section \ref{sec:properties}. In this section, we endow the state space with a partial order and show that the optimal value function is monotone decreasing in state. We further establish the Lipschitz continuity of the value function in this section. Thereafter, we turn our attention to proving the main result -- the convergence of fitted value iteration -- in Section \ref{sec:proof}. Having computed the approximate value function for the aggregate scheduling, we identify the individual EV scheduling algorithm in Section \ref{sec:schedule}. The numerical simulations are presented in Section \ref{sec:numerical}. Finally, we conclude the paper in Section \ref{sec:conclusion}.  

\subsection{Notations}
$\mathbb{O}^{d\times d'}$ is an $d\times d'$ dimensional all zero matrix, $\mathbb{I}^{d\times d}$ is an $d\times d$ dimensional identity matrix. Let $\CAL X$ be a set. In this paper, $\CAL C_b(\CAL X)$ denotes the space of continuous and bounded function endowed with supremum norm on the set $\CAL X$.

\section{Problem Formulation}\label{sec:scheduling}
We formulate the EV charge scheduling problem as a dynamic optimization problem with a finite time horizon $\CAL T=\{1,...,T\}$. Consider an operator that provides a menu-based charging service. The menu here is denoted by $\CAL B\subset\{1,\ldots,M\}\times\{1,\ldots,N\}$, where an item $(m,n)\in\CAL B$ means that the facility will provide $m$ units of electricity within $n$ time slots. For instance, suppose that $\CAL B = \{10\kwh,20\kwh,...,50\kwh\}\times\{1\h,2\h,...,8\h\}$. If an EV arrives at the charging station at $8$ AM and leaves at $3$ PM demanding $37 \kwh$ electricity, then it will choose the item $(m,n) = (40 \kwh,7 \h)$. We define $\CAL B_n = \{m:(m,n)\in \CAL B\}$. For simplicity and ease of exposition, we assume that all EVs charge with a constant charging rate given by $r \text{ kW}$. 

The operator assigns a category $(t,m,n)\in\CAL T\times\CAL B$ to every EV depending on the preferences of the EV owner: Here, $t\in\CAL T$ is the arrival time of the EV and $(m,n)\in\CAL B$ is the item selected by the EV owner. The goal of the charge scheduling problem is to allocate electricity to each category $(t,m,n)$ at each time $s\in\CAL T$ with minimal total operating costs. Let $\CAL I_s\subset\CAL T\times\CAL B$ denote the categories of EVs that are present at time $s$: This is given by $\CAL I_s = \{(t,m,n): t\leq s\leq t+n-1\}$. Define $\CAL I^1_s = \{(t,m,n): s = t+n-1 \}$ to be the categories of the EVs that are connected at time $s$ but will depart at $s+1$ and $\CAL I^2_s = \CAL I_s\setminus \CAL I^1_s$.

Let $w^{t,m,n}$ be the number of EVs selecting item $(m,n)$ that arrive at time $t$. This is a non-negative integer valued random variable with a known distribution. We assume that $w^{t,m,n}$ is bounded, that is, $w^{t,m,n}\in\{0,\ldots,\bar w^{t,m,n}\}$ for some $\bar w^{t,m,n}<\infty$. We let $w^{t,m,n}=0$ whenever $t< 0$. We further assume that the sequence of random variables $\{w^{t,m,n}\}_{t\in\CAL T,(m,n)\in\CAL B}$ are mutually independent. Let $w_t$ be the random vector representing all new arrivals at time $t$:
\begin{align*}
    w_t=&\Big[(w^{t,m,1})_{m\in\CAL B_1},\ldots,(w^{t,m,N})_{m\in\CAL B_N}\Big]^\transpose \in\CAL W_t:=\prod_{(m,n)\in\CAL B}\{0,\ldots,\bar w^{t,m,n}\}\subset\Na^{\abs{\CAL B}}.
\end{align*}
We let $y_s$ denote the vector of number of EVs at the charging station in each category in $\CAL I_s$:
\begin{align*}
    y_s:=(w^{t,m,n})_{(t,m,n)\in\CAL I_s}\in\CAL Y_s:=\prod_{(t,m,n)\in\CAL I_s}\{0,\ldots,\bar w^{t,m,n}\},
\end{align*}
where $y_s$ is formed in the order of leaving time, i.e.
\begin{align*}
    y_s=\Big[& \underbrace{(w^{s-1,m,1})_{m\in\CAL B_1},...,(w^{s-N,m,N})_{m\in\CAL B_N}}_{\text{leaving at }s}, ..., \underbrace{0,...,(w^{s-1,m,N})_{m\in\CAL B_N}}_{\text{leaving at }s+N-1}\Big].
\end{align*}

At each time $s$, the total electricity allocated to the EVs in the category $(t,m,n)\in \CAL I_s$ is denoted by $u_s^{t,m,n}$. We let $u_s$ be the vector of the electricity allocation to each category $(t,m,n)\in\CAL I_s$:
\begin{align*}
    u_s = (u^{t,m,n}_s)_{(t,m,n)\in\CAL I_s}\in\CAL U_s := \Re_+^{\abs{\CAL I_s}}.
\end{align*}
We assume that for $t<0$, we let $u_s^{t,m,n}=0$. We also have the constraint that the total electricity allocated to all the categories be in the interval $[d^1_s,d^2_s]$, that is, $ d^1_s\leq \mathds{1}^\transpose u_s \leq d^2_s$. Let $d_s = (d^1_s,d^2_s)^\transpose$.

Suppose that allocating one unit of electricity to $(t,m,n)$ at time $s$ incurs a cost $c_s^{t,m,n}$. Then, the total cost to the operator at each time is $c_s^\transpose u_s$, where 
\begin{align*}
    c_s := (c^{t,m,n}_s)_{(t,m,n)\in\CAL I_s}\in\Re^{\abs{\CAL I_s}}.
\end{align*}
The notion of the cost here is general: either this could be the cost of electricity in the wholesale electricity market or it could be the cost of electricity minus the valuation of the electricity to the EV owner. Accordingly, we assume that the cost can be positive or negative.

\subsection{State of the System}
Let $z_s^{t,m,n}$ be the remaining electricity required by the category $(t,m,n)\in\CAL I_s$, defined as
\begin{align*}
    z_s^{t,m,n} = 
    \begin{cases}
    my^{t,m,n} & s = t\\
    my^{t,m,n} - \sum_{\tau = s-t}^{s-1}u_s^{t,m,n} & t\leq s\leq t+n-1\\
    0 & s \geq t+n
    \end{cases},
\end{align*}
It is easy to see that the update equation for $z_s^{t,m,n}$ is given by
\begin{align}\label{eq:z}
    z_{s+1}^{t,m,n} = 
    \begin{cases}
    my^{t,m,n} & s = t-1\\
    z_s^{t,m,n} - u_s^{t,m,n} & t\leq s\leq t+n-2\\
    0 & s \geq t+n-1
    \end{cases},
\end{align}
Define $z_s$ as $(z^{t,m,n}_s)_{(t,m,n)\in\CAL I_s}$ and let $\CAL Z_s\subset \Re_+^{\abs{\CAL I_s}}$ be the set of all such $z_s$. 

We now define $x_s = [y_s^\transpose,z_s^\transpose,d_s^\transpose]^\transpose$ as the state of the system at time $s$, and $\CAL X_s := \CAL Y_s\times\CAL Z_s\times \Re_+^2$ is the corresponding state space of the system. Further, $u_s$ is the action of the system. For each state $x_s\in\CAL X$, the feasible action $u_s$ should satisfy that, at each time $s$, the allocated electricity is upper bounded by minimum of the remaining electricity $z_s$ and the charging capacity $ry_s$. Let $g:\CAL X_s\to\CAL U_s$ be given by $g(x_s):=\min\{ry_s, z_s\}$, where $\min\{a,b\}:=[\min\{a_i,b_i\}]_i$ is the elementwise minimum of two vectors $a,b$. Thus, the feasible action set at state $x_s$ is defined by a correspondence $\Gamma:\CAL X_s\rightrightarrows \CAL U_s$, which is given by
\begin{align}\label{eq:cons}
    \Gamma(x_s):=&\Big\{u_s\in\CAL U:\;  0\leq u_s\leq g(x_s),d^1_s\leq \mathds{1}^\transpose u_s \leq d^2_s\nonumber\\
    &\qquad u^{t,m,n}_s = z^{t,m,n}_s \text{ for all } (t,m,n)\in\CAL I^1_s\Big\}.
\end{align}
Let $\CAL D_s$ be the joint feasible state-action pairs, defined as
\beqq{\CAL D_s :=\{(x,u)\in\CAL X_s\times\CAL U_s: u\in\Gamma(x_s)\}.}
In the next subsection, we determine the transition dynamics of the state of the system. 
\subsection{State Dynamics and Admissible Policies}
The system is a linear dynamical, where the state transition function is $f$, which is given by
\begin{align}\label{eq:update}
    &x_{s+1} = f(x_s,u_s,w_s,d_{s+1}):=
    \begin{bmatrix}
    A_y y_s + C_y w_s\\
    A_z (z_s-u_s) + C_z w_s\\ d_{s+1}
    \end{bmatrix} =\nonumber\\
    &  
    \underbrace{\begin{bmatrix} A_y & 0 & 0\\ 0& A_z & 0\\ 0 & 0 & 0\end{bmatrix}}_{=A} x_s + \underbrace{\begin{bmatrix} 0\\-A_z\\ 0\end{bmatrix}}_{=B} u_s + \begin{bmatrix} C_y \\ C_z\\0\end{bmatrix}w_s+\begin{bmatrix} 0 \\ 0\\\mathbb{I}^{2\times 2}\end{bmatrix}d_{s+1},
\end{align}
where the time invariant matrices $A_y$, $A_z$, $C_y$, and $C_z$ are given as follows:
\begin{align*}
    A_y=& A_z =
    \begin{bmatrix}
        \mathbb{O}^{\abs{\CAL I^2_s}\times \abs{\CAL I^1_s}} & \mathbb{I}^{\abs{\CAL I^2_s}\times \abs{\CAL I^2_s}}\\
          \mathbb{O}^{\abs{\CAL I^1_s}\times \abs{\CAL I^1_s}}&   \mathbb{O}^{\abs{\CAL I^1_s}\times \abs{\CAL I^2_s}}
    \end{bmatrix}, \nonumber\\ 
    C_y=&
    \begin{bmatrix}
      C^1 & 0 & ... & 0 \\
      0 & C^2 & ... & 0 \\
      \vdots & \vdots & & \vdots\\
      0 & 0 & ... & C^N
    \end{bmatrix}, \
    C^k=
    \begin{bmatrix}
      0\\  \vdots \\ \BBM{I}^{\abs{\CAL B_k}\times \abs{\CAL B_k}} \\ \vdots \\ 0
    \end{bmatrix},
    \nonumber\\ 
    C_z^k =&
    \begin{bmatrix}
      0\\  \vdots \\ \text{diag}\Big(\{m\}_{m\in\CAL B_k}\Big) \\ \vdots \\ 0
    \end{bmatrix},\ 
    \begin{matrix}
    \text{where}\\
    \CAL B_k = \{(m,n)\in\CAL B: n = k\}.
    \end{matrix}
\end{align*}
In the model above, one can consider $d_{s+1}$ to be a deterministic ``actuation noise'' whose distribution is a Dirac mass at the point $d_{s+1}$.


At time $s$, a feasible policy is a measurable map $\pi_s:\CAL X_s\to\CAL U_s$ such that $\pi_s(x_s)\in\Gamma(x_s)$ for all $x_s\in\CAL X_s$. Let $\Pi_s$ denote the set of all feasible policies. We further let $\pi=(\pi_0,...,\pi_T)$ denote a feasible strategy of the operator and $\Pi:=\prod_{s=0}^T \Pi_s$ denote the feasible strategy space.

\subsection{Performance Index and Dynamic Programming Based Optimal Solution}
We are now ready to introduce the finite time horizon stochastic dynamic programming (DP). The expected total cost of the system given the initial state $x$ using the strategy $\pi$ is given by  
\begin{align*}
   J(\pi;x) = \cex{\sum_{s=1}^T c_s^\transpose \pi_s(x_s)}{x_1=x},
\end{align*}
The goal is to minimize the expected total cost from $s=1$ to $s=T$ given the initial state $x$. This problem can be solved using the usual dynamic programming method under fairly mild conditions. The optimal value functions $v_s$ can be obtained by applying Bellman operator $H_s$ for each time $s=1,...,T$, which is
\begin{align}\label{eq:v_s}
    v^*_s(x_s)=H_{s+1}(v^*_{s+1})(x_s):=\inf_{u_s\in\Gamma(x_s)} c_s^\transpose u_s +\ex{v^*_{s+1}\left(f(x_s,u_s,W_s,d_{s+1})\right)},
\end{align}
where we take $v^*_{T+1}(x_{T+1}) \equiv 0$.

In this case, we can obtain the optimal scheduling policies $\pi^*=[\pi_s^*]_{s\in\CAL T}$ by applying value iteration $v^*_s=H_{s+1}(v^*_{s+1})$ in \eqref{eq:v_s} from time $s=T$ to time $s=1$ recursively. Here, $\pi_s^*(x_s)$ is the minimizer in \eqref{eq:v_s}.

\subsection{Key Challenges}
When $\abs{\CAL B}$ is sufficiently large, the dimensionality of the state space and action space is also large. In this case, computing the $v_s$ for each $s\in\CAL T$ is significantly challenging due to the curse of dimensionality. Thus, we propose to use approximate dynamic programming to compute approximately optimal value functions.

In this problem, the state transition function, the cost function, the demand distribution, and the state-dependent action set are known {\it a priori}. Thus, we have only two major hurdles to overcome for computing the approximately optimal value functions: the computation of expectation in the Bellman operator is very challenging and the approximately optimal value function needs to be stored in computers for future use. To alleviate the first challenge, we propose to use the empirical Bellman operator, which uses i.i.d. samples of noise to approximate the computation of the expected future value. To alleviate the second challenge, we use a projection operator that takes as input the values from the computation of the empirical Bellman operator and outputs a function in the chosen function approximating class. These techniques are pretty standard in the reinforcement learning literature. The algorithm and the corresponding convergence result are described in the next section.

\section{Empirical Bellman Operator and Fitted Value Iteration}\label{sec:efvi}
We use empirical Bellman operator $\hat H_{s+1}^k:\CAL C_b(\CAL X_s)\to\CAL C_b(\CAL X_s)$ to approximate the actual Bellman operator $H_{s+1}$. Let $\{W_{s,i}\}_{i=1}^k$ be a sequence of independent identically distributed (i.i,d.) samples of $w_s$, then the empirical Bellman operator $\hat H_{s+1}^k$ is given by
\begin{align}\label{eq:emp}
   \hat v^k_s(x_s)&=\hat H_{s+1}^k(\hat v^k_{s+1})(x_s):=\inf_{u_s\in\Gamma(x_s)} c_s^\transpose u_s +\frac{1}{k}\sum_{i=1}^k \hat v^k_{s+1}\left(f(x_s,u_s,W_{s,i})\right).
\end{align}

\subsection{Fitted Value Iteration}\label{sec:fvi}
While applying the value iteration, it is necessary to store a function approximator of $\hat v^k_s$ in computers. The function approximator can be obtained by projecting the value function $\hat v^k_s$ onto a feasible function approximating class, such as neural networks or reproducing kernel Hilbert space (RKHS), which is dense in $\CAL C_b(\CAL X)$.

Consider a function approximating class $\CAL G_d(\CAL X_s)\subset\CAL C_b(\CAL X_s)$, whose size is parameterized by $d\in\Na$. For instance, $d$ can represent the number of perceptrons in a neural network or the number of basis functions in RKHS. We create a dataset for each time $s\in\CAL T$, i.e. $\{x_{s,j},\hat v^k_s(x_{s,j})\}_{j=1}^l$, where $\{x_{s,j}\}_{j=1}^l$ are uniformly sampled from the state space $\CAL X_s$, and $\hat v^k_s(x_{s,j})$ is obtained according to \eqref{eq:emp}. We consider a loss function $\text{Loss}:\CAL C_b(\CAL X_s)\times\CAL G_d(\CAL X_s)\to\Re_+$ that measures the distance between $v_s\in\CAL C_b(\CAL X_s)$ and $h\in\CAL G_d$. A common loss function can be picked as the mean squared error:
\begin{align*}
    \text{Loss}\left(\hat v^k_s,h\mid\{x_{s,j}\}_{j=1}^l\right)=\frac{1}{l}\sum_{j=1}^l\left(\hat v^k_s(x_{s,j})-h(x_{s,j})\right)^2.
\end{align*}
We denote $\Pi_s^{l,d}:\CAL C_b(\CAL X_s)\to\CAL G_d(\CAL X_s)$ as the function approximating projection that maps the output of $\hat H_{s+1}^k(\hat v^k_{s+1})$ to a function in $\CAL G_d$. This is defined as
\begin{align}\label{eq:Pi}
    \Pi_s^{l,d}(\hat v^k_s)=\arg\inf_{h\in\CAL G_d}\text{Loss}\left(\hat v^k_s,h\mid\{x_{s,j}\}_{j=1}^l\right).
\end{align}

\subsection{Composition of Random Operators}
We here construct a composited operators that combines the empirical Bellman operator and function approximating operator. We let
\begin{align*}
    \Psi_s^{k,l,d}=\Pi_s^{l,d} \circ \hat H_{s+1}^k: \CAL G_d(\CAL X_{s+1}) \to \CAL G_d(\CAL X_s)
\end{align*}
be the random fitted empirical Bellman operator used in place of the actual Bellman operator $H_{s+1}$ to arrive at an approximate function $\hat v_s$. Here, $k$ is the number of samples generated, $d$ is a parameter describing the size of the function approximating class, and $l$ is the number of samples used in computing the empirical loss function for the projection operation.

We define the \textit{fitted value iteration} at time $s\in\CAL T$ as
\begin{align*}
     \hat v_s^{k,l,d}(x_s)=\Psi_s^{k,l,d}\Big(  \hat v_{s+1}^{k,l,d}\Big)(x_s).
\end{align*}
We now proceed to proving that this fitted value iteration algorithm converges as we increase $k,l,d\to\infty$. In what follows, we aim at increasing the $k,l,d$ simultaneously. Let $j\in\Na$ and $k(j), l(j),d(j)$ be such that as $j\to\infty$, we have $k(j), l(j),d(j)\to\infty$. By a slight abuse of notation, we denote $\hat H^j_{s+1} := \hat H^{k(j)}_{s+1}$, $\Pi^j_s := \Pi^{l(j),d(j)}_s$, and the fitted value iteration algorithm by 
\begin{align*}
     \hat v_s^j(x_s)=&\Psi_s^j\left(\hat v_{s+1}^j\right)(x_s) :=\Psi_s^{k(j),l(j),d(j)}\left(  \hat v_{s+1}^{k(j),l(j),d(j)}\right)(x_s).
\end{align*}
To establish the convergence of the proposed algorithms, we also need the following reasonable assumptions on the projection operators.

\begin{assumption}\label{asp:proj_v}
The projection operator $\Pi^{l,d}_s :\CAL C_b(\CAL X_s)\to\CAL G_d(\CAL X_s)$ satisfies the followings two conditions:
\begin{renum}
    \item $\Pi^{l,d}_s$ is approximately non-expansive, that is, for all $v_1, v_2 \in \CAL C_b(\CAL X_s)$, we have 
    \begin{align*}
        \norm{\Pi^{l,d}_s(v_1) - \Pi^{l,d}_s(v_2)}_\infty \leq \norm{v_1 - v_2}_\infty + \hat{\zeta}^{l,d}_s,
    \end{align*}
    where $\hat{\zeta}^{l,d}_s\leq \bar\zeta_s<\infty$ almost surely and $\hat{\zeta}^{l,d}_s\to 0$ as $l,d\to\infty$ in probability.
    \item For any $\epsilon > 0$ and $\delta > 0$, there exists $M_l,M_d$ that may depends on $v^*_s$ such that
    \begin{align*}
        \pr{\norm{\Pi^{l,d}_s(v^*_s) - v^*_s}_\infty > \epsilon } < \delta \text{ for all } l \geq M_l, d\geq M_d.
    \end{align*}
\end{renum}
\end{assumption}
Under the assumptions listed above, we have the following theorem where the convergence of the fitted value iteration algorithm is established.

\begin{theorem} \label{thm:pfv}
If Assumption \ref{asp:proj_v} holds, then $\hat v_s^j$ satisfies for any $\kappa > 0$,
\begin{align*}
    \limsup_{j \rightarrow \infty} \pr{\norm{\hat v_s^j - v^*_s}_\infty > \kappa } = 0.
\end{align*}
\end{theorem}
The proof of the Theorem is established in Section \ref{sec:proof}. Thus, as we increase the number of samples for empirical Bellman operator, expand the function approximating class to include more parameters, and take more samples of the state to project the value function to the function approximating class, we are guaranteed to converge to the optimal value functions under the sup norm.

\section{Properties of Value Functions}\label{sec:properties}
In this section, we study three crucial properties of the value function -- monotonicity and Lipschitz continuity with respect to the state $x_s$, and continuity with respect to the system parameters.

\subsection{Monotonicity of Value Functions}
Note that any realization of the state $x_s$ is a non-negative vector in $\Na^{\abs{\CAL I_s}}\times\Re_+^{\abs{\CAL I_s}}$.  Endow the state space $\CAL X_s$ with the following partial order: Let $x_s,x_s'\in\CAL X$. Then, $x_s\leq x_s'$ if and only if $y_s\leq y_s'$, $z_s^{t,m,n}= z'^{t,m,n}_s$ for every $(t,m,n)\in\CAL I^1_s$, $z^{t,m,n}_s\leq z'^{t,m,n}_s$ for every $(t,m,n)\in\CAL I^2_s$, and $d_s\leq d_s'$. A function $v:\CAL X_s\to\Re$ is said to be a {\it monotonically increasing} function if and only if for any $x,x'\in\CAL X_s$ such that $x\leq x'$, we have $v(x)\leq v(x')$. A function $v:\CAL X\to\Re$ is said to be a {\it monotonically decreasing} function if and only if $-v$ is monotonically increasing. In this section, we show that the dynamic optimization problem formulated above yields \textit{monotonically decreasing} value functions at all times.

\begin{theorem}\label{thm:monotone}
For each $s\in\CAL T$, the optimal value function $v^*_s$ is a monotonically decreasing function of $x_s$.
\end{theorem}
\begin{proof}
To show this, we first note that for any $x\leq x'$, we have
\begin{renum}
    \item $\Gamma(x) \subseteq \Gamma(x')$.
    \item $f(x, u, w,d_{s+1}) \leq f(x', u, w,d_{s+1})$ for all $u \in \Gamma(x)$ and $w\in\CAL W_s$.
\end{renum}
We now prove the statement using induction. The terminal cost is 0, so it is trivially monotone decreasing. Assume that $v_{s+1}$ is monotonically decreasing. We claim that $v^*_s = H_{s+1}(v^*_{s+1})$ is also monotone decreasing function. Pick $x,x'\in\CAL X_s$ such that $x\leq x'$, $u\in\Gamma(x)$ and $w\in\CAL W_s$. Since $f(x,u,w)\leq  f(x',u,w)$ and $v^*_{s+1}$ is monotonically decreasing, we conclude that
\begin{align}\label{eq:vf}
    v^*_{s+1}(f(x', u, w,d_{s+1}))\leq v^*_{s+1}(f(x, u, w,d_{s+1})) .
\end{align}

Consequently, $\ex{v^*_{s+1}(f(\cdot, u, W,d_{s+1})} $ is also monotonically decreasing function. This yields 
\begin{align*}
    & \inf_{u\in\Gamma(x)}c_s^\transpose u + \ex{v^*_{s+1}(f(x, u, W,d_{s+1}))} \\
    &\qquad \geq \inf_{u\in\Gamma(x')} c_s^\transpose u + \ex{v^*_{s+1}(f(x, u, W,d_{s+1}))}\\
    &\qquad \geq \inf_{u\in\Gamma(x')} c_s^\transpose u + \ex{v^*_{s+1}(f(x', u, W,d_{s+1}))},
\end{align*}
where the first inequality is due to $\Gamma(x) \subseteq \Gamma(x')$, and the second inequality results from \eqref{eq:vf}.

In other words, $v^*_s$ is monotonically decreasing. An application of the principle of mathematical induction implies that $v^*_s$ is monotone decreasing for all $s$.
\end{proof}

\subsection{Lipschitz Continuity of Value Functions}
We now endow the state and the action space with metrics and establish the Lipschitz continuity of the value functions. Let $\CAL X:=\CAL X_0 = \CAL X_2 =\ldots =\CAL X_T$ and a same convention is applied for $\CAL U$. Define the metric on $\CAL X$ and $\CAL U$ as 
\begin{align*}
    \rho_X(x,x')=\norm{x-x'}_\infty, \quad \rho_U(u,u')=\norm{u-u'}_\infty,
\end{align*}
for any $x,x'\in\CAL X, u,u'\in\CAL U$. Let $2^{\CAL U}$ denote the set of all compact subsets of $\CAL U$. We endow this space with the Hausdorff metric, given by 
\begin{align*}
    \rho_H(\SF U,\SF U') = \max \Big\{ \sup_{u\in \SF U} \inf_{u' \in \SF U'} \rho_U(u, u'), \sup_{u'\in \SF U'} \inf_{u \in \SF U} \rho_U(u, u')\Big\},
\end{align*}
for all $\SF U,\SF U'\subset \CAL U$. 
\begin{theorem}\label{thm:cts}
The value function $v^*_s$ is a Lipschitz continuous function.
\end{theorem}
\begin{proof}
We first claim the following statements:
\begin{renum}
    \item The correspondence $\Gamma: \CAL X_s \rightarrow 2^{\CAL U_s}$ is Lipschitz continuous with coefficient $L_\Gamma = \max\{r,1\}$: For any $x, x' \in \CAL X_s$, we have
    \begin{align*}
        \rho_H (\Gamma(x), \Gamma(x')) \leq L_{\Gamma} \rho_X(x, x').
    \end{align*}
    \item For every $w \in \CAL W_s$, the state transition function $f(\cdot, \cdot, w)$ is Lipschitz continuous in $(x, u) \in \CAL D_s$ with Lipschitz coefficient  $L_f(w) \equiv 1$ and $L_P := \int L_f(w) \pr{dw} = 1 < \infty$.
    \item The cost function $c_s:\CAL D_s \rightarrow \Re$ is Lipschitz continuous with Lipschitz coefficient $L_{c_s}:=\|c_s\|_1$.
\end{renum}
We can write $\Gamma(x)$ as $\Gamma(x) = \{u\in\CAL U_s: u\geq 0, Q_1 u \leq Q_2 x, Q_3 u = Q_4 x \}$ for appropriate matrices $Q_1,Q_2,Q_3,Q_4$ that have bounded entries. Thus, the constraint set is actually a polyhedral set. From the discussion in Section 1 of \cite{li1994sharp}, we conclude that $\Gamma$ is a Lipschitz continuous correspondence with Lipschitz coefficient $L_\Gamma$. The exact value of Lipschitz coefficient is difficult to derive, and we refer the reader to \cite{li1994sharp} and \cite{walkup1969lipschitzian} for a detailed discussion on upper bounds on $L_\Gamma$.

We now prove the second claim. Using triangle inequality, we have
\begin{align*}
    \norm{f(x,u,w)-f(x',u',w)}_\infty & \leq \norm{A}_\infty\norm{x-x'}_\infty + \norm{B}_\infty\norm{u-u'}_\infty \\
    & \leq \left(\norm{x-x'}_\infty + \norm{u-u'}_\infty\right)
\end{align*}
which shows that $f$ is Lipschitz continuous over $\CAL D_s$ with Lipschitz coefficient $1$. The Lipschitz coefficient of the cost function is derived from the Cauchy Schwarz inequality.

The Lipschitz continuity of the value function then follows from Theorem 4.1 in \cite[p. 12]{hinderer2005lipschitz}. We present an outline here. Suppose that $v^*_{s+1}$ is Lipschitz continuous with Lipschitz coefficient $L_{v^*_{s+1}}$. Then, we use Theorem 4.1 in \cite[p. 12]{hinderer2005lipschitz} to conclude that
\begin{align*}
    \abs{v^*_s(x_s)-v^*_s(x'_s)} &\leq \abs{ c_s^\transpose u_s^* -c_s^\transpose u'^*_s } +\Big| \ex{v^*_{s+1}(f(x_s,u_s^*,w_s,d_{s+1}))}\\
    &\qquad -\ex{v^*_{s+1}(f(x_s',u'^*_s,w_s,d_{s+1}))}\Big| \\
    &\leq L_{c_s}L_\Gamma \norm{x_s-x_s'}_\infty + L_{v^*_{s+1}}(1+ L_{\Gamma})\norm{x_s-x_s'}_\infty\nonumber\\
    &\leq (L_{c_s}L_\Gamma+L_{v^*_{s+1}}L_P(1+L_\Gamma))\norm{x_s-x_s'}_\infty,
\end{align*}
which implies $v^*_s$ is Lipschitz continuous with Lipschitz constant $L_{v^*_s} = L_{c_s}L_\Gamma+L_{v^*_{s+1}}(1+L_\Gamma)$ (since $L_P =1$). The induction step is complete.
\end{proof}

\subsection{Robustness of Value Functions with respect to Parameters}
Our problem here has multiple parameters that can change over time. For instance, the cost of acquiring electricity in the wholesale markets or the distribution of the EV arrival process may change slightly over time. This can be studied under the umbrella of parameterized dynamic programs, where the parameters influence the cost/profit functions or the EV arrival process. We investigate in this section the continuity of the value function as a function of the parameters. We identify some sufficient conditions under which a slight change in the parameters would lead to a slight change in the value function. This allows us to conclude the robustness of the scheduling algorithm with respect to small parametric uncertainty.

Let $\Theta\subset\Re^q$ be the parameter space, which is assumed to be a compact subset of a Euclidean space. We consider a parameterized optimization problem, parameterized by $\theta\in\Theta$, in which 
\begin{enumerate}
    \item $\tilde c_s(\theta)$ is the negative profit function;
    \item The probability distribution of the EV arrival process $\tilde W_s$ is given by $\nu_s(\cdot,\theta)$.
\end{enumerate}

The parameterized dynamic program is then rewritten as:
\begin{align*}
    \tilde v_s^*(x_s,\theta)= \inf_{u_s\in\Gamma(x_s)}  \tilde c_s(\theta)^\transpose u_s +\exsub{\nu_s(\theta)}{\tilde v^*_{s+1}\left(f(x_s,u_s,\tilde W_s,d_{s+1}),\theta\right)}.
\end{align*}
Here, $\tilde v_s^*:\CAL X_s\times\Theta\to\Re$ is the optimal parameterized value function. We also let $\tilde \pi_s^*(x_s,\theta)$ be the corresponding parameterized scheduling policy. We identify some sufficient conditions and establish the continuity of $\tilde v_s^*$ and lower semicontinuity of $\tilde \pi_s^*$ below.

\begin{assumption}\label{asm:para_cts}
The following holds
\begin{renum}
    \item $\tilde c_s$ is continuous on $\Theta$;
    \item There exists a base probability measure $\lambda_s$ and a continuous and bounded function $\beta_s:\CAL W_s\times\Theta\to[0,\infty)$ such that 
\begin{align*}
    \nu_s(dw,\theta) = \beta_s(w,\theta) \lambda_s(dw).
\end{align*}
\end{renum}
\end{assumption}

\begin{theorem}\label{thm:cts_c}
Suppose that Assumption \ref{asm:para_cts} holds. Then, $\tilde v^*$ is jointly continuous on $\CAL X_s\times\Theta$ and $\tilde \pi_s^*$ is lower semi-continuous on $X_s\times\Theta$. 
\end{theorem}
\begin{proof}
We apply the result from \cite[Theorem 1]{dutta1994parametric} to establish this result. First, Assumption \ref{asm:para_cts} (i) implies the cost function $(u_s,\theta)\mapsto\tilde c_s(\theta)^\transpose u_s$ is jointly continuous on $\Theta\times\CAL U_s$. 

Recall that the state transition function $f$ is a linear map (see \eqref{eq:update}). Then linearity of $f$ and Assumption \ref{asm:para_cts}(ii) implies that for any $h\in\CAL C_b(\CAL X_{s+1})$ and any convergent sequence $\{(x_n,u_n,\theta_n)\}_n\subset \CAL X_s\times\CAL U_s\times\Theta$ satisfying $(x_n,u_n,\theta_n)\to(x,u,\theta)$, we have $h(f(x_n,u_n,w,d))\beta_s(w,\theta_n)\to h(f(x,u,w,d))\beta_s(w,\theta)$. Further, since $h,\beta_s$ are continuous and bounded functions, we conclude that
\begin{align*}
    &\lim_{n\to\infty}\int h(f(x_n,u_n,w,d'))\nu_s(dw,\theta_n)\\
    =&\lim_{n\to\infty}\int h(f(x_n,u_n,w,d'))\beta_s(w,\theta_n)\lambda_s(dw)\\
    \stackrel{(a)}{=}&\int h(f(x,u,w,d'))\beta_s(w,\theta)\lambda_s(dw)\\
    =& \int h(f(x,u,w,d'))\nu_s(dw,\theta),
\end{align*}
where the equality in (a) results from the dominated convergence theorem as $\CAL X_s,\CAL U_s,\CAL W_s,\Theta$ are compact.

Note that we have also shown in the proof of Theorem \ref{thm:cts} (statement (i)) that $\Gamma(x_s)$ is a continuous and compact-valued correspondence. Thus, an application of \cite[Theorem 1]{dutta1994parametric} implies that $\tilde v_s^*$ is continuous on $\CAL X_s\times\Theta$ and $\tilde \pi_s^*$ is lower semi-continuous on $X_s\times\Theta$ , which completes the proof.




\end{proof}





\section{Proof of Theorem \ref{thm:pfv}}\label{sec:proof}
We first establish two auxiliary results to establish the theorem. The first statement establishes that the empirical Bellman operator is nonexpansive. The second statement shows that the empirical Bellman operator $\hat H^j_{s+1}$ when applied on $v^*_{s+1}$ converges to $v^*_s$ in probability as $j\to\infty$. 

\begin{lemma}\label{lem:Hnonexp}
For any $v,v'\in\CAL C_b(\CAL X_{s+1})$ and any realization of the random operator $\hat H^j_{s+1}$, we have
\beqq{\|\hat H^j_{s+1}(v) - \hat H^j_{s+1}(v')\|_\infty \leq \|v-v'\|_\infty \text{ almost surely}.}
\end{lemma}
\begin{proof}
The proof is straightforward and therefore omitted. 
\end{proof}

\begin{lemma}\label{lem:lipf}
For any $\epsilon > 0$, we have the following holds:
\begin{align*}
    \lim_{k\rightarrow \infty} & \pr{\Big\|\hat{H}^k_{s+1}(v^*_{s+1}) -H_{s+1}(v^*_{s+1})\Big\|_\infty \geq \epsilon } = 0,
\end{align*}
\end{lemma}
\begin{proof}
The proof is in \ref{app:lipf}.
\end{proof}
We now proceed to proving Theorem \ref{thm:pfv} using the principle of mathematical induction. We have
\begin{align}\label{eqn:hatvjs}
    \norm{\hat v^j_s - v^*_s}_\infty\leq  \norm{\Psi^j_s(\hat v^j_{s+1}) - \Psi^j_s(v^*_{s+1})}_\infty  +\norm{\Psi^j_s(v^*_{s+1}) -H_{s+1}(v^*_{s+1})}_\infty.
\end{align}
Let us consider the first summand on the right side of the equation above. We have 
\begin{align*}
    \norm{\Psi^j_s(\hat v^j_{s+1}) - \Psi^j_s(v^*_{s+1})}_\infty  &\leq \norm{\hat H^j_{s+1}(\hat v^j_{s+1}) - \hat H^j_{s+1}(v^*_{s+1})}_\infty + \zeta^j_s\\
    & \leq \norm{\hat v^j_{s+1} - v^*_{s+1}}_\infty + \zeta^j_s,
\end{align*}
where we used Lemma \ref{lem:Hnonexp} and Assumption \ref{asp:proj_v}(i). Next, consider the second summand on the right side of \eqref{eqn:hatvjs}:
\begin{align*}
     \norm{\Psi^j_s(v^*_{s+1}) -H_{s+1}(v^*_{s+1})}_\infty &=\norm{\Pi^j_s(\hat H^j_{s+1}(v^*_{s+1})) -v^*_s}_\infty\\
    & \leq\norm{\Pi^j_s(\hat H^j_{s+1}(v^*_{s+1})) -\Pi^j_s(v^*_s)}_\infty + \norm{\Pi^j_s(v^*_s) -v^*_s}_\infty\\ 
    & \leq\norm{\hat H^j_{s+1}(v^*_{s+1})) -v^*_s}_\infty + \zeta^j_s + \norm{\Pi^j_s(v^*_s) -v^*_s}_\infty,
\end{align*}
where the first inequality is due to the triangle inequality and the second inequality is due to Assumption \ref{asp:proj_v}(i).
Thus, we conclude that 
\begin{align*}
    &\norm{\hat v^j_s - v^*_s}_\infty\leq  \\ &\qquad \norm{\hat v^j_{s+1} - v^*_{s+1}}_\infty + \norm{\hat H^j_{s+1}(v^*_{s+1})) -v^*_s}_\infty + \norm{\Pi^j_s(v^*_s) -v^*_s}_\infty + 2\zeta^j_s.
\end{align*}
For time $s = T$, we have $v^*_{T+1} =\hat v^j_{T+1} = 0$. As $j\to\infty$, all three terms on the right goes to 0 in probability due to Lemma \ref{lem:lipf}, Assumption \ref{asp:proj_v}(i), and Assumption \ref{asp:proj_v}(ii). Thus, $\norm{\hat v^j_T - v^*_T}_\infty\to 0$ in probability as $j\to\infty$ and the statement holds for time $T$.

For any time $s$, we can use the same argument to conclude that as $j\to\infty$, $\norm{\hat v^j_s - v^*_s}_\infty\to\infty$ in probability. The proof of the theorem is complete.

\section{Scheduling Using a Forward DP}\label{sec:schedule}
Having computed the value functions using the fitted value iteration algorithm described above, we now need to determine a method to compute $u_1$ using the state $x_1$, $u_2$ using the state $x_2$, etc. We now introduce the forward DP algorithm that calculates the optimal charging policy. Note that if we do not charge the EV with enough amount of electricity at the beginning, then we may not satisfy the total requirement at the end of the charging process due to the charging rate constraint. In other words, the feasible action set $\Gamma(x_s)$ in \eqref{eq:cons} may be empty at time $s$ if $u_1,...,u_{s-1}$ were not picked appropriately. For instance, at time $s$, the equality constraint $z_s^{t,m,n}=u_s^{t,m,n}$ for $(t,m,n)\in\CAL I^1_s$ can become incompatible with the inequality constraint $0\leq u_s\leq g(x_s)$ whenever $u_1,...,u_{s-1}$ are not large enough.

\begin{remark}
The above problem -- in which the state dependent action set $\Gamma(x_s)$ may become empty -- arises because we are computing an approximate value function $\hat v^j_{s+1}$. If we could compute $v^*_{s+1}$ exactly and use this for determining $u^*_s$ (using the usual dynamic programming recursion), then this problem would not arise and the state dependent action set would be nonempty at all times.
\end{remark}

To address this problem, we add additional constraints  that iteratively guarantees $\Gamma(x_s)$ to be non-empty: We denote $\Gamma'(x_s)$ as the new feasible action set with 
\begin{align}\label{eq:gamma'}
    \Gamma'(x_s)&=\Gamma(x_s)\cap \nonumber\\
    &\ \left\{u_s\in\CAL U: z_{s+1}^{t,m,n}\leq r y_s^{t,m,n}(t+n-s-2),\text{ for all } (t,m,n)\in\CAL I_s^2\right\},
\end{align}
where $z_{s+1}^{t,m,n}$ is updated according to \eqref{eq:z}. Here, $r y_s^{t,m,n}(t+n-s-2)$ represents the maximum amount of electricity can be charged in the remaining charging window, which should be larger than the remaining demand $z_{s+1}^{t,m,n}$. Using this modified state dependent constraint, we compute the approximately optimal action as
\begin{align}\label{eqn:hatus}
    u^*_s = \hat\pi_s(x_s) :=\underset{u_s\in\Gamma'(x_s)}{\arg\min} c_s^\transpose u_s +\frac{1}{k}\sum_{i=1}^k \hat v^j_{s+1}\left(f(x_s,u_s,W_{s,i})\right).
\end{align}

This guarantees $u_t^{t,m,n},..., u_{s-1}^{t,m,n}$ are large enough so that $z_s^{t,m,n}\leq g(x_s)^{t,m,n}$ for $(t,m,n)\in\CAL I^1_s$. Thus, the inequality in \eqref{eq:gamma'} ensures that if $\Gamma'(x_s)$ is non-empty, then $\Gamma'(x_{s+1}),\Gamma'(x_{s+2}),...$ are also non-empty. Moreover, the set $\Gamma'(x_s)$ is deterministic (depending on the current state $x_s$ only) even if the system dynamics \eqref{eq:update} involves stochastic arrivals of EVs in the future.

From the forward DP, we can obtain the optimal amount of electricity allocated to each category ${u_s^{t,m,n}}^*$. Then each individual EV in the category can be charged according to a simple disaggregation algorithm: each EV will receive $r$ amount of electricity for ${{u_s^{t,m,n}}^*}/{(ry_s^{t,m,n})}$ fraction of the time in the time interval from $s$ to $s+1$. The overall algorithm for determining the EV charging algorithm is as follows in Algorithm \ref{alg:c}.

\begin{algorithm}
\caption{Preemptive Scheduling Algorithm of EV Charging}
\begin{algorithmic}\label{alg:c}
\STATE \textbf{Part I: Fitted Value Iteration}
\STATE Initialize $v_{T+1}^*\equiv 0$.
\FOR{$s=T,...,1$}
    \STATE Generate the state and noise samples $\{x_{s,j}\}_{j=1}^l$ and $\{W_{s,i}\}_{i=1}^k$. 
    \STATE Create data set $\{x_{s,j},\hat v^k_s(x_{s,j})\}_{j=1}^l$ using \eqref{eq:emp} and obtain $\hat v_s^j$ according to \eqref{eq:Pi}.
\ENDFOR
\STATE \textbf{Part II: Forward DP}
\STATE Initialize $x_1=0$.
\FOR{$s=1,...,T$}
    \STATE Update $x_s$ using \eqref{eq:update} and $\Gamma'(x_s)$ using \eqref{eq:gamma'}.
    \STATE Compute $u_s^*$ with \eqref{eqn:hatus}.
    \FOR{$(t,m,n)\in \CAL T\times\CAL B$}
    \STATE Charge each EV in category $(t,m,n)$ for $\frac{{u_s^{t,m,n}}^*}{ry_s^{t,m,n}}\in[0,1]$ fraction of the time in the interval from $s$ to $s+1$. 
    \ENDFOR
\ENDFOR
\end{algorithmic} 
\end{algorithm}

\section{Numerical Results}\label{sec:numerical}
In this section, we show a series of numerical simulations to demonstrate the performance of our ADP algorithm.

\subsection{Simulation Setup}
We consider the scheduling of EV charging processes for a $T=24$ hour period, that is, from 7 AM (day 1) to 7 AM (day 2). The electricity prices vary according to the peak/off-peak hours. The customers pay a constant price $9.2\cent/\text{kWh}$ for charging their EVs. The cost $c_s$ is considered as the difference between the electricity price and the charging revenue from the customers. The details are shown in Table \ref{tab:price}. Note that here, $c_s$ is interpreted as the negative of the profit.

\begin{table}[!ht]
\centering
\setlength{\tabcolsep}{3pt}
\begin{threeparttable}
\caption{Electricity prices during the weekdays \label{tab:price}}
\begin{tabular}{c c c c c}
    \toprule
    Time (h) & 7-14 & 15-18 & 19-22 & 23-7 \\
    \hline
    Peak hours & Mid-Peak & On-Peak & Mid-Peak & Off Peak \\
    Grid prices (\cent/\kwh) & 9.2 & 16.6 & 9.2 & 4.8 \\
    $c_s$ (\cent/\kwh) & 0 & 7.4 & 0 & -4.4 \\
    \bottomrule
\end{tabular}
\end{threeparttable}
\end{table}

\begin{figure}
\centering
\includegraphics[width=\linewidth]{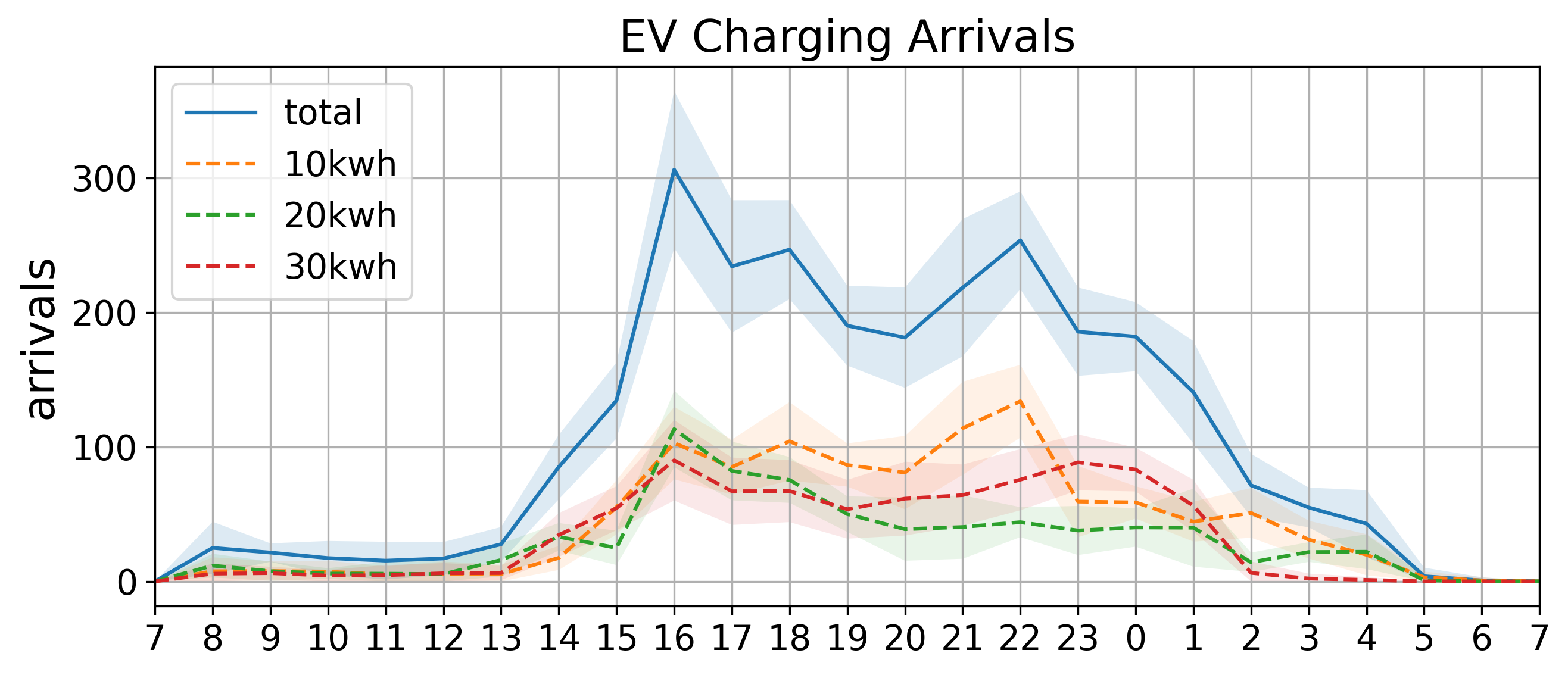}
\caption{The arrivals of EVs at each time. The solid line represents the total arrivals over the $10$ sample paths at each time $s\in\CAL T$. The dashed lines are the sum of arrivals at each time with demands in $\{10\kwh,20\kwh,30\kwh\}$ respectively.}
\label{fig:dem}
\end{figure}

We pick $M=3$ and $N=6$, and the charging rate is fixed at $r = 10 \text{ kW}$. We also pick $d_s=[0 \kwh,10000 \kwh]$ as the hourly grid bounds. The feasible menu $\CAL B$ is shown in Table \ref{tab:menu} with $\abs{\CAL B}=15$. With this menu, the dimensionality of the state $x_s$ is $\text{dim}(\CAL X_s)=182$ and the action is $\text{dim}(\CAL U_s)=90$. The arrival process $\{w_t\}_{t\in\CAL T}$ is a sequence of random variables with Poisson distribution, where the parameters (means) are chosen according to the ACN-Data \cite{lee_acndata_2019} shown in Figure \ref{fig:dem}.

\begin{table}[!ht]
\centering
\setlength{\tabcolsep}{0.5pt}
\begin{threeparttable}
\caption{Feasible menu given $M=3$ and $N=6$ \label{tab:menu}}
\begin{tabular}{ m{0.2\textwidth}>{\centering\arraybackslash}m{0.12\textwidth}>{\centering}m{0.12\textwidth}>{\centering\arraybackslash}m{0.12\textwidth}>{\centering\arraybackslash}m{0.12\textwidth}>{\centering\arraybackslash}m{0.12\textwidth}>{\centering\arraybackslash}m{0.12\textwidth} }
    \toprule
    $\CAL B$ & n=1\h & n=2\h & n=3\h & n=4\h & n=5\h & n=6\h\\ \hline
    m=10\kwh & (1,1) & (1,2) & (1,3) & (1,4) & (1,5) & (1,6) \\ 
    m=20\kwh &  \xmark & (2,2) & (2,3) & (2,4) & (2,5) & (2,6) \\ 
    m=30\kwh & \xmark & \xmark & (3,3) & (3,4) & (3,5) & (3,6) \\
    \bottomrule
\end{tabular}
\end{threeparttable}
\end{table}

For the projection operator, we pick the number of state samples $l=64$ and the number of noise samples $k=64$. The function approximating class $\CAL G_d$ is the set of neural networks with width $\abs{\CAL X_s}\times 2=364$ and depth $8$. During the training process of the neural network, the learning rate is chosen as $0.005$. The EDP algorithm is employed to compute the value functions $\hat v^j_s$.

\subsection{Results}
We demonstrate the performance of our ADP algorithm, denoted as ADP, by comparing it with two other algorithms: SP and FCFS. Algorithm SP solves the optimal action with the full knowledge of all the future demands. In this case, the problem can be formulated by a \textit{static} (or deterministic) program since all the realization of $\{w_t\}_{t\in\CAL T}$ is known. We denote the optimal actions of SP as $\{u_{\text{SP},s}^*\}_{s\in\CAL T}$. The second algorithm FCFS follows the First Come First Serve discipline, which charges the EVs immediately when they arrive at the charging station. This is the most widely used scheduling algorithm across the world. Let the actions of FCFS be denoted by $\{u_{\text{FCFS},s}^*\}_{s\in\CAL T}$. Table \ref{tab:alg} provides the information required by the three algorithms discussed in this paper.

\begin{table}[!ht]
\centering
\caption{Application scenarios of the algorithms given knowledge of the future \label{tab:alg}}
\begin{tabular}{c c c c}
    \toprule
    Algorithms & future demands & demand distribution & no knowledge \\
    \hline
    SP & \cmark & \xmark & \xmark \\ 
    ADP & \cmark & \cmark & \xmark\\ 
    FCFS & \cmark & \cmark & \cmark \\
    \bottomrule
\end{tabular}
\end{table}

\begin{figure}
\centering
\includegraphics[width=\linewidth]{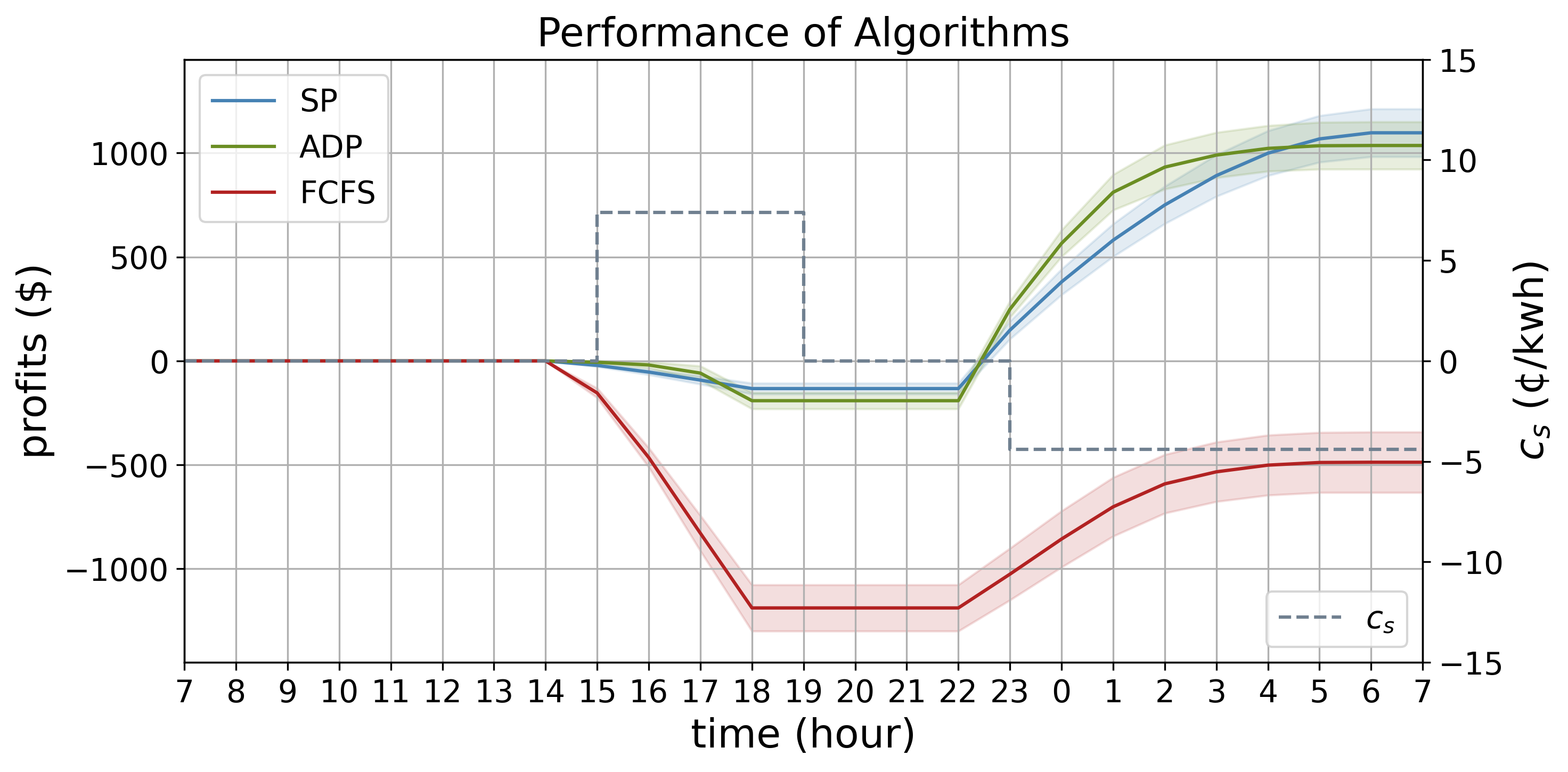}
\caption{Performance of ADP, SP, and FCFS with 10 sample paths. For better understanding, we plot the cumulative profits (negative cumulative cost, $-J_{\alpha,t}^*$) verse time $t$ under different algorithms. The right axis represents the $\{c_s\}_{s\in\CAL T}$. The ADP achieves a similar profit with SP, and a higher profit in comparison with FCFS at the end of the time horizon. }
\label{fig:cc}
\end{figure}

\begin{figure}
\centering
\includegraphics[width=\linewidth]{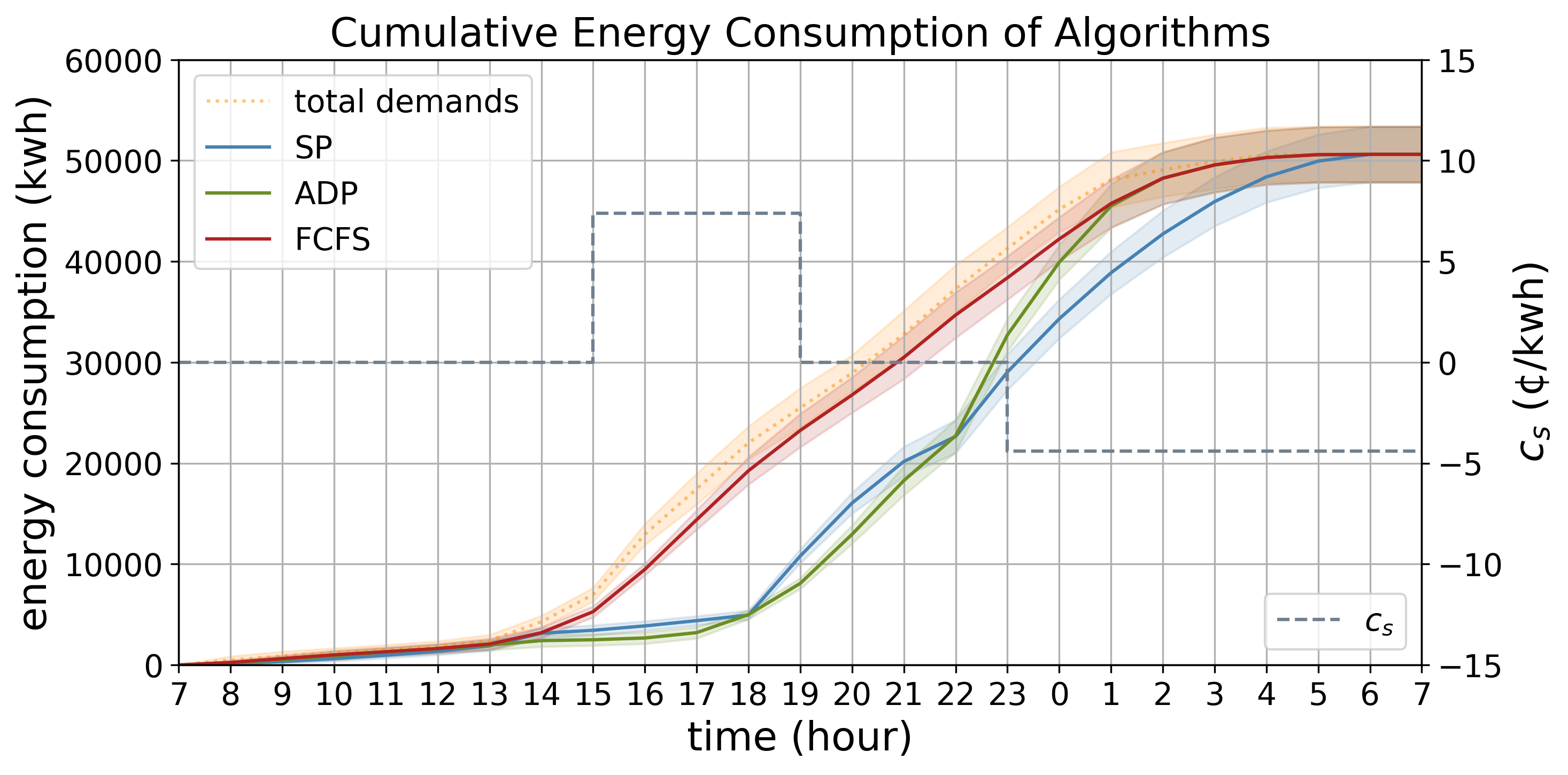}
\caption{The cumulative energy consumption $\sum_{s=1}^t \mathds{1}^\transpose u_{\alpha,s}^*$ given $10$ sample paths. All algorithms complete the demands at the end of the time horizon, whereas, the curves of SP and ADP increase much slower than FCFS during the time slots with high $c_s$, which indicates that these two algorithms postpone the charging processes during the peak hours in order to increase profits.}
\label{fig:ca}
\end{figure}

We let the optimal actions of our ADP algorithm be $\{u_{\text{ADP},s}^*\}_{s\in\CAL T}$. To compare the performance, we let 
\begin{align}\label{eq:J_sim}
    J_{\alpha,t}^* = \sum_{s=1}^t c_s^\transpose u_{\alpha,s}^*,\  \alpha\in\{\text{ADP},\text{SP},\text{FCFS}\}
\end{align}
be the cumulative costs for each sample path. Note that $J_{\text{SP}}^*$ provides lower bounds on $J_{\text{ADP}}^*$ and $J_{\text{FCFS}}^*$ since it knows all the future demand. We compare $J_{\alpha,t}^*$ of these three algorithms over $10$ sample path of the noises $\{w_t\}_{t\in\CAL T}$ in Figure \ref{fig:cc}. We see that since ADP and SP exploit knowledge of future demand distribution or demand itself, it leads to much higher profit in comparison to FCFS charging policy. We also notice that there is very little loss in performance while using ADP, which has knowledge of only future demand distribution, in comparison to using SP, which has knowledge of the entire future demand. This may be attributed to the distributional assumption on the EV arrival process $\{w_t\}_{t\in\CAL T}$, which was deduced from the ACN-Data \cite{lee_acndata_2019}. To have a detailed comparison, we also plot the cumulative energy consumption respectively in Figure \ref{fig:ca}.

\begin{figure}
\centering
\includegraphics[width=\linewidth]{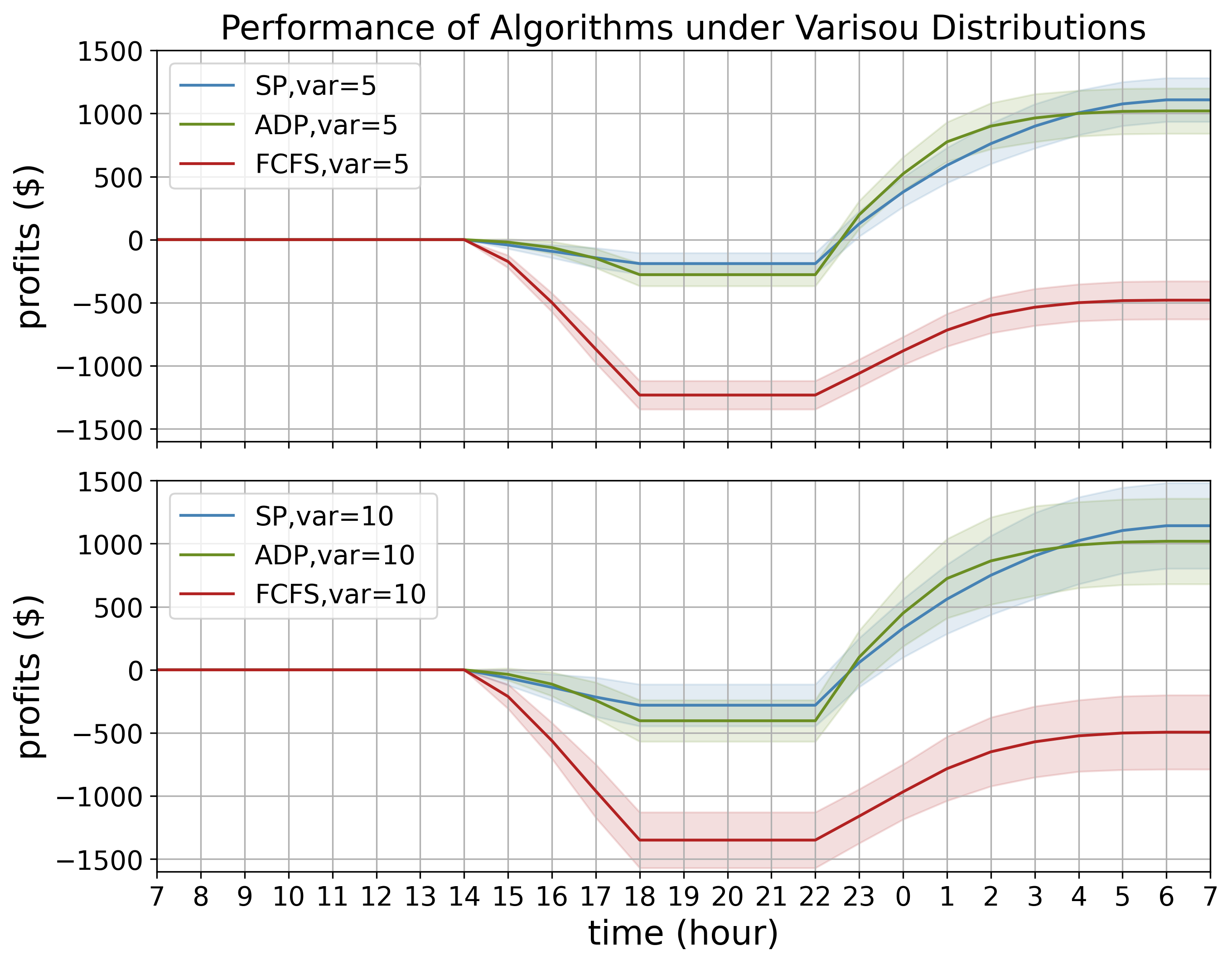}
\caption{Cumulative profits of ADP, SP and FCFS where $\{w_t\}_{t\in\CAL T}$ are chosen as truncated Gaussian random variable with $\text{Var}=5$ (top) and $\text{Var}=10$ (bottom). The variance of the $J_{\alpha,t}^*$ increases for a larger variance of $\{w_t\}_{t\in\CAL T}$. }
\label{fig:cc_var}
\end{figure}

We also simulate the profits when the arrivals $\{w_t\}_{t\in\CAL T}$ are distributed with the truncated Gaussian distribution. The means of $\{w_t\}_{t\in\CAL T}$ remain the same as in the previous setting, and the variances are chosen from $\{5,10\}$. The profits ($-J^*_{\alpha,t}$) under different variances are shown in Figure \ref{fig:cc_var}. It yields that the cumulative costs $J_{\text{ADP},T}^*$ remains near-optimal even if the distributions of the $\{w_t\}_{t\in\CAL T}$ are perturbed slightly, which verifies the results in Theorem \ref{thm:cts_c}.

\begin{figure}
\centering
\includegraphics[width=\linewidth]{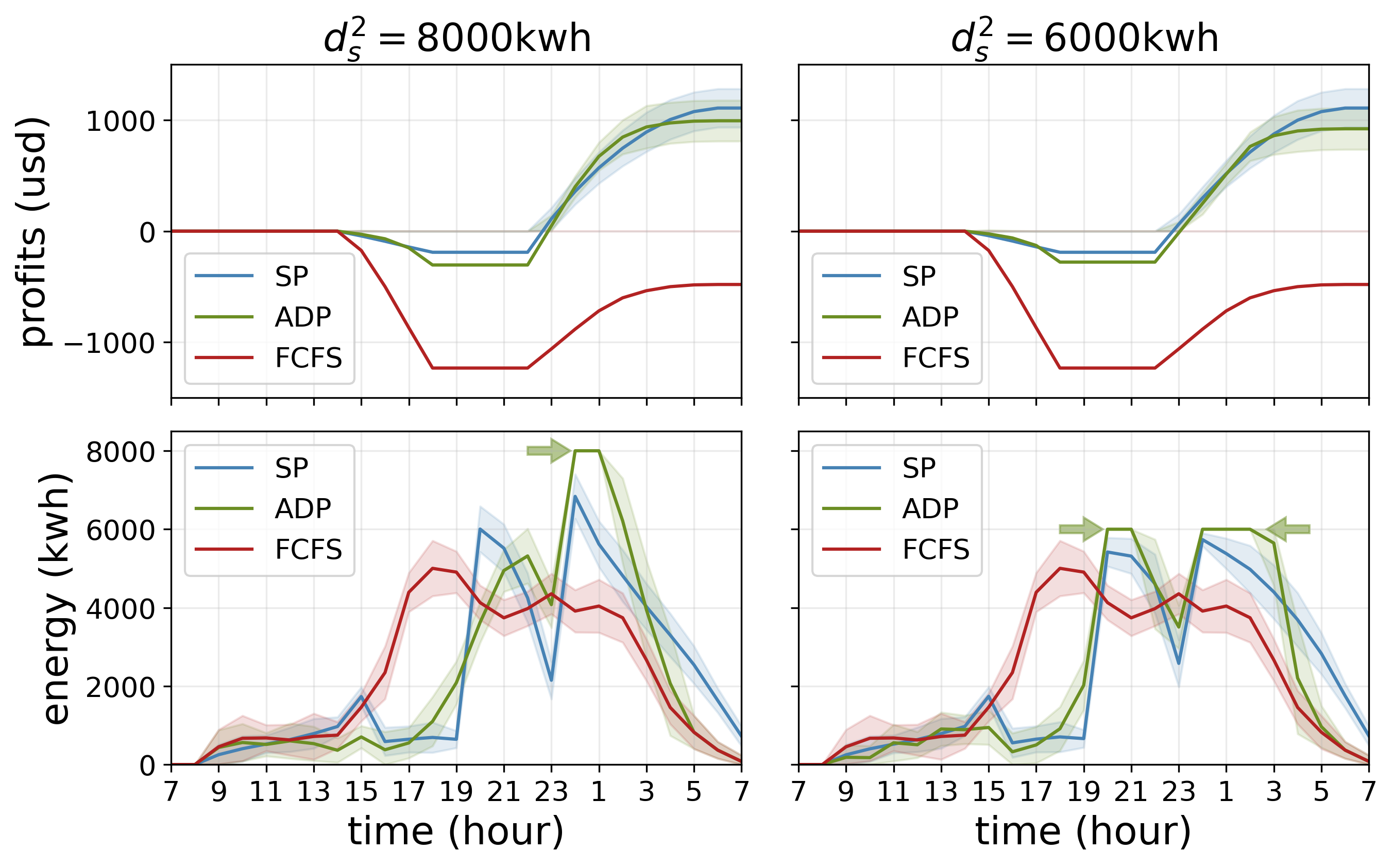}
\caption{Cumulative profits of ADP, SP and FCFS given $d_s^2=8000\kwh$ (left), and $d_s^2=6000\kwh$ (right). The upper figures plot the cumulative profits, whereas the lower figure plot the energy consumption for each hour. The green arrows point the time when the constraint $\mathds{1}^\transpose u_s \leq d^2_s$ in \eqref{eq:cons} is active. ADP is not able to allocate enough energy to charge EVs, which leads to a lower profit at the end of the time horizon.}
\label{fig:cc_d}
\end{figure}

We finally demonstrate our algorithm performance under different grid bounds $d_s$. In this case we pick $d_s^2\in\{8000\kwh,6000\kwh\}$ while keeping $d_s^1=0\kwh$. Note that we need to ensure the existence of the feasible solution for SP, that is, we need $d_s^2-d_s^1$ large enough to guarantee $\Gamma(x_s)\neq\emptyset$ by \eqref{eq:cons}. The results are shown in Figure \ref{fig:cc_d}. ADP has a worse performance under a lower grid bound $d_s^2$ since the approximation error to the value function $v_s^*$, especially for the state $x_s\in\CAL X_s$ with $\Gamma(x_s)$ that the inequality constraint $\mathds{1}^\transpose u_s \leq d^2_s$ in \eqref{eq:cons} is active. 

\section{Conclusion}\label{sec:conclusion}
In this paper, we modeled the problem of scheduling a large number of electric vehicles in a smart grid as a stochastic dynamic program. We showed that the value function satisfies a monotonicity property and is Lipschitz continuous, and demonstrate it is robust against perturbation of the system parameters. We proposed to use fitted value iteration to solve the problem due to the very high dimensional state and action spaces. We exploited the Lipschitz continuity of the value functions to show the consistency of the fitted value iteration algorithm. Our simulations shows that our algorithm yields profits that are close to the optimal profits under full information about the future demand of the EVs. For the simulations, we assumed that the EVs arrive according to the distribution that was inferred from the ACN-Data \cite{lee_acndata_2019}. We also demonstrated through simulations that our scheduling algorithm is robust to perturbations in the distributional assumptions on the EV charging demand.


\appendix
\section{Proof of Lemma \ref{lem:lipf}}\label{app:lipf}
We need to introduce some notations. 
For $(x,u) \in \CAL D_s$, let $\xi_{x,u}: \CAL W_s\times\Re_+^2 \rightarrow \Re$ be defined as $\xi_{x,u}(w,d_{s+1}) = v^*_{s+1}(f(x,u,w,d_{s+1}))$. Let $\Xi$ be a collection of such functions as $ \Xi = \bigcup_{(x,u) \in \CAL D_s} \{\xi_{x,u}(\cdot)\}$.
Note that we have 
\begin{align*}
    &\norm{\hat{H}^k_{s+1}(v^*_{s+1}) - H_{s+1}(v^*_{s+1})}_\infty \\
    = & \sup_{x\in\CAL X_s} \Bigg| \inf_{u\in \Gamma(x_s)} \left(c_s^\transpose u + \frac{1}{k} \sum_{i=1}^{k}v^*_{s+1}(f(x,u, W_{s,i},d_{s+1})) \right)\\
    &\qquad - \inf_{u\in \Gamma(x_s)} \left(c_s^\transpose u +\int v^*_{s+1}(f(s,a, w,d_{s+1}))\pr{dw} \right)\Bigg|\\
    \leq & \sup_{(x,u)\in \CAL D_s} \abs{ \frac{1}{k}\sum_{i=1}^{k} \xi_{x,u}(W_{s,i}) - \int \xi_{x,u}(w) \pr{dw} }\\
    = & \sup_{\xi \in \Xi}  \abs{ \frac{1}{k}\sum_{i=1}^{k} \xi(W_{s,i}) - \int \xi(w) \pr{dw} }.
\end{align*}
Then, for any $\epsilon>0$, we conclude
\begin{align} \label{ieq:brktn}
   &\pr{\norm{\hat H^k_{s+1}(v^*_{s+1}) - H_{s+1}(v^*_{s+1})}_\infty \geq \epsilon }  \nonumber\\ 
   &\qquad \leq \pr{ \sup_{\xi \in \Xi}\abs{ \frac{1}{k}\sum_{i=1}^{k} \xi(W_{s,i}) - \int \xi(w) \pr{dw} }
   \geq \epsilon}.
\end{align}
To show the right side of (\ref{ieq:brktn}) converges to $0$ as $k$ goes to infinity, we show that the bracketing number of $\Xi$ is finite for each $\epsilon > 0$. Since $v^*_{s+1}(\cdot)$ and $f(\cdot, \cdot, w,d_{s+1})$ are Lipschitz continuous function for every $w \in \CAL W_s$ and $d_{s+1}$, for all $(x, u), (x', u') \in \CAL D_s$, we have
\begin{align*}
    \abs{ \xi_{x,u}(w) - \xi_{x',u'}(w) } \leq L_{v^*_{s+1}} \rho_D ((x,u),(x',u')).
\end{align*}
According to \cite[Theorem 2.7.11]{van1996weak}, the bracketing number of $\Xi$ is bounded by the covering number of $\CAL D_s$. Since $\CAL X_s\times \CAL U_s$ is compact and $\Gamma(\cdot)$ is Lipschitz continuous, $\CAL D_s$ is a compact set. Thus, the bracketing number of $\Xi$ is bounded for each $\epsilon>0$. Then by~\cite[Theorem 2.4.1]{van1996weak}, we conclude that the right side of (\ref{ieq:brktn}) converges to $0$ as $k$ goes to infinity.

\bibliographystyle{plain}
\bibliography{reference,prob,probbook,math}

\end{document}